\documentclass[sigconf]{acmart}

\newif\ifARXIV

\ARXIVtrue

\usepackage{prettyref}

\newrefformat{eqn}{Equation~\ref{#1}}
\newrefformat{fig}{Figure~\ref{#1}}
\newrefformat{tab}{Table~\ref{#1}}
\newrefformat{sec}{Section~\ref{#1}}
\newrefformat{lst}{Listing~\ref{#1}}
\newrefformat{alg}{Algorithm~\ref{#1}}
\def\BibTeX{{\rm B\kern-.05em{\sc i\kern-.025em b}\kern-.08emT\kern-.1667em\lower.7ex\hbox{E}\kern-.125emX}}

\ifARXIV

\setcopyright{none}
\renewcommand\footnotetextcopyrightpermission[1]{} 
\else

\copyrightyear{2023}
\acmYear{2023}
\setcopyright{acmlicensed}\acmConference[CF '23]{20th ACM International Conference on Computing Frontiers}{May 9--11, 2023}{Bologna, Italy}
\acmBooktitle{20th ACM International Conference on Computing Frontiers (CF'23), May 9--11, 2023, Bologna, Italy}
\acmPrice{15.00}
\acmDOI{10.1145/3587135.3592172}
\acmISBN{979-8-4007-0140-5/23/05}

\fi

\author{G.~Palermo, G.~Accordi, D.~Gadioli,  E~.Vitali, C.~Silvano, B.~Guindani, D.~Ardagna}
\affiliation{%
  \institution{DEIB - Politecnico di Milano}
  \city{Milan}
  \country{Italy}
}

\author{A.~R.~Beccari, D.~Bonanni, C.~Talarico, F.~Lughini, }
\affiliation{
  \institution{Dompé Farmaceutici Spa}
  \city{Napoli}
  \country{Italy}
}

\author{J.~Martinovic, P.~Silva, A.~Bohm, J.~Beranek, J.~Krenek, B.~Jansik}
\affiliation{
  \institution{IT4Innovations, VSB – TUO}
  \city{Ostrava}
  \country{Czech Republic}
}

\author{L.~Crisci, B.~Cosenza}
\affiliation{
  \institution{University of Salerno}
  \city{Fisciano (SA)}
  \country{Italy}
}

\author{P.~Thoman, P.~Salzmann, T.~Fahringer}
\affiliation{
  \institution{University of Innsbruck}
  \city{Innsbruck}
  \country{Austria}
}

\author{L. T. Alexander, G. Tauriello, T. Schwede, J. Durairaj}
\affiliation{
  \institution{Biozentrum, University of Basel, }
\institution{SIB Swiss Institute of Bioinformatics}
  \city{Basel}
  \country{Switzerland}
}

\author{A.~Emerson, F.~Ficarelli}
\affiliation{
  \institution{CINECA}
  \city{Bologna}
  \country{Italy}
}

\author{S.~Wingberm\"uhle, E.~Lindahl}
\affiliation{
  \institution{KTH Royal Institute of Technology}
  \city{Stockholm}
  \country{Sweden}
}

\author{D.~Gregori}
\affiliation{
  \institution{E4 Computer Engineering Spa}
  \city{Scandiano (RE)}
  \country{Italy}
}

\author{E.~Sana, S.~Coletti}
\affiliation{
  \institution{Chelonia}
  \city{Lugano}
  \country{Switzerland}
}

\author{P.~Gschwandtner}
\affiliation{
  \institution{PH3 GmbH, University of Innsbruck}
  \city{Innsbruck}
  \country{Austria}
}

\begin{document}

%
\title{Tunable and Portable Extreme-Scale Drug Discovery Platform at Exascale: the LIGATE Approach}

%


\renewcommand{\shortauthors}{G. Palermo, et al.}

%
\begin{abstract}
Today digital revolution is having a dramatic impact on the pharmaceutical industry and the entire healthcare system.
The implementation of machine learning, extreme-scale computer simulations, and big data analytics in the drug design and development process offers an excellent opportunity to lower the risk of investment and reduce the time to patient.

Within the LIGATE project \footnote{LIGATE website - \url{https://www.ligateproject.eu/}}, we aim to integrate, extend, and co-design best-in-class European components to design Computer-Aided Drug Design (CADD) solutions exploiting today's high-end supercomputers and tomorrow's Exascale resources, fostering European competitiveness in the field.

The proposed LIGATE solution is a fully integrated workflow that enables to deliver the result of a virtual screening campaign for drug discovery with the highest speed along with the highest accuracy.
The full automation of the solution and the possibility to run it on multiple supercomputing centers at once permit to run an extreme scale in silico drug discovery campaign in few days to respond promptly for example to a worldwide pandemic crisis.

%


\end{abstract}

%
%
\ifARXIV
\else 
\fi
%


%

%
\maketitle

\section{Introduction}

The pharmaceutical industry is among the sectors investing in R\&D more than any other. So far most of the investments in R\&D were not in computational tools, but rather in wet chemistry, and clinical trials. In recent years, the situation changed with growing investments in computational tools (HPC in particular) to reduce the cost for the development of new drugs or shorten the time to market and digital transformation in general. The investment in HPC, besides becoming fundamental for competitiveness, is being recognized as having a high ROI (Return of Investment), for the pharmaceutical sector. 
HPC has been used in the field of drug discovery to accelerate the process of finding new drugs since it permits researchers to analyze large chemical libraries against selected targets in a short amount of time, allowing them to identify potential treatments more quickly. 
Indeed, the embarrassingly parallel nature of the virtual screening process \cite{murugan2022review} makes it suitable for exploiting large computing infrastructures composed of powerful compute nodes.

In this direction, the LIGATE project supports the development of a European drug-discovery platform capable to exploit recent HPC investments, guaranteeing independence from other economic areas in the world, and reducing the risk of suffering from digital gaps with respect to competitors.

The LIGATE computer-aided drug discovery (CADD) platform will be designed with the following key characteristics:
\begin{itemize}
    \item[(i)] \emph{Portability}, to execute on all the pre-exascale and future exascale European supercomputers; 
    \item[(ii)] \emph{Tunability}, to select the optimal configuration depending on the target goal of small-/extreme-scale virtual screening campaign;
    \item[(iii)] \emph{Scalability}, to run on all available resources dedicated to the experiment, from a single heterogeneous node to a multi-site execution in case of urgent computing.
\end{itemize}

LIGATE is a three-year project co-funded by the European High-Performance Computing (EuroHPC) Joint Undertaking under the topic \emph{Industrial software codes for extreme scale computing environments and applications}. It is led by Dompè Farmaceutici (Project Coordinator) and Politecnico di Milano, and aggregates overall 10 partners among Universities, Research Institutions, and Companies, including CINECA, VSB-IT4Innovation, E4 Computer Engineering, University of Salerno, University of Innsbruck, KTH - Royal Institute of Technology, University of Basel, PH3, and Chelonia.

The remainder of the document is organized as follows: Section~2 briefly describes the LIGATE Computer-Aided Drug Discovery workflow.
Section~3 lists and describes the main project goals by outlining the main achievement reached so far.
Finally, Section~4 concludes the paper.

\section{LIGATE CADD Workflow}

\begin{figure}[t]
    \centering
    \includegraphics[width=\linewidth]{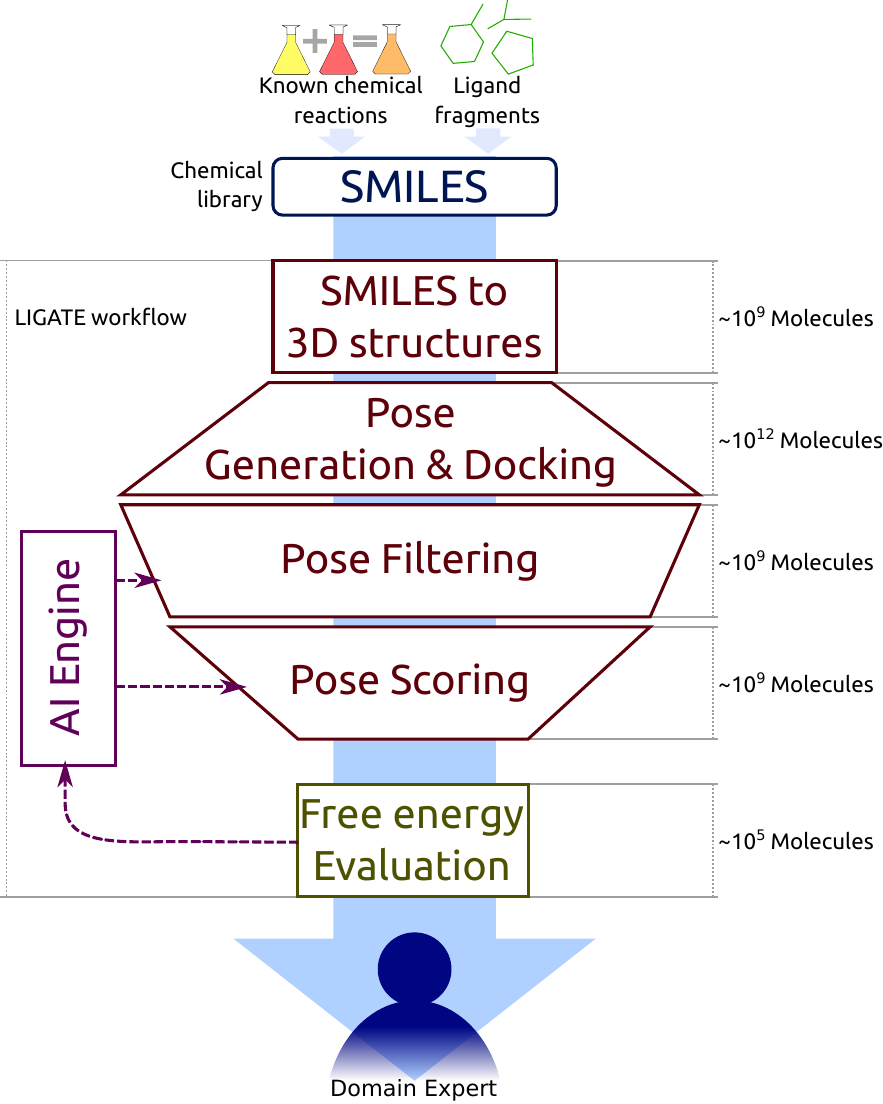}
    \caption{Overview of the LIGATE workflow including also the number of molecules that it is expected to analyze at each stage during an extreme-scale virtual screening campaign}
    \label{fig:LIGATEWF}
\end{figure}

The main goal of a virtual screening campaign is to suggest to a domain expert a relatively small set of promising molecules,
that have been predicted with the strongest interaction (high-affinity values) with at least one binding site of the target proteins. 
This output will drive the selection of molecules that will proceed to the \textit{in-vitro} and \textit{in-vivo} stages of the classical drug discovery process.
Recent studies show how the introduction of this \textit{in-silico} phase increases the success probability \cite{matter2011application, allegretti2022repurposing}. 

\prettyref{fig:LIGATEWF} provides an overview of the LIGATE CADD workflow.
The proposed workflow is a structure-based approach since it uses the 3D shape of both the ligands and the target protein as the basis for the evaluation. 
While we explain all the components in the order required to virtually screen a chemical library, its implementation is modular.
This enables us to create on-demand additional workflows by arranging the required components in a different way.

The chemical library that we want to analyze is composed of virtual molecules that are derived by simulating known chemical reactions or by composing known ligand fragments.
In this way, we have access to a very large chemical space.
To limit its storage requirement, we use the SMILES format, which encodes a molecule in a single text line.
Moreover, we use a custom algorithm to compress/decompress each line independently using a single dictionary that is independent of the input file.
The second kind of input is the target proteins, which are required by almost all the components of the CADD workflow, encoded using the PDB format.
We are agnostic with respect to the methodology used to resolve them \cite{ijms21145152}.

The SMILES notation of a molecule is compact because it describes only the topological shape of the molecule.
The first component of the workflow aims at computing all the information required by the other ones, such as the 3D displacement of the molecule's atoms.
Then, we dock the ligand in the target binding sites.
The adopted docking algorithm uses a gradient descent with multiple restarts that promote diversity in the generated poses, increasing the chemical library size.
The docking gradient is a simple scoring function that takes into account only geometric information.
Therefore, we use an additional component to re-score the generated poses, taking into account also physico-chemical information.
We consider the ligand score equal to the one of its best pose.
Since the computation effort required by the latter component can have a big impact on the performance, we can have an intermediate step between them.
In particular, the pose filtering step aims at discarding the poses that are not likely to yield a high score.
These four components are implemented using LiGen. 

LiGen is a high throughput virtual screening software part of the EXSCALATE platform \footnote{EXSCALATE website - \url{https://exscalate.com/}} that can scale up to a full modern supercomputer. It is a computational toolbox for molecule manipulation having its core in the dock and score phase.
It is used to quickly estimate the interaction strength between a protein and a small molecule, and then for selecting a relatively small set of promising candidates. 
In the context of the EXCALATE4CoV European Project, LiGen virtually screened more than $70$-billion ligands docking and scoring them in $15$ binding sites of $12$ viral proteins of SARS-CoV-2 \cite{9817028}. The experiment lasted $60$ hours, and it involved two supercomputers with a combined throughput peak of $81$ PFLOPS.

Once the scoring function has been used to evaluate the input chemical library, we can rank them to select which molecule we want to investigate further by evaluating the absolute binding free energy (ABFE).
The estimation of the affinity at which a small molecule binds to a protein (ABFE) happens by studying an all-atom molecular dynamics (MD) of the complex in a solution.
This component requires the highest computational effort among the available ones.
Therefore, we can't evaluate all the molecules in the initial chemical library, but we rely on the previous stages to select the most promising candidates and to provide a reasonable initial displacement of the molecule's atoms to start the free energy evaluation.
The domain experts can use this output to support the selection of the molecules to test in-vitro.
This component is implemented using the GROMACS software package.

GROMACS provides a fast and highly portable MD implementation that efficiently uses both MPI and OpenMP parallelization.  GROMACS can offload nearly all computations of an MD step to the GPU using CUDA, OpenCL, and SYCL. Its kernels have been optimized for 11 CPU SIMD instruction sets. GROMACS also features a variety of enhanced-sampling techniques, i.e., algorithms that reduce the number of MD steps needed to achieve a precise binding affinity estimate.

While the actual LIGATE CADD pipeline ends after the free energy evaluation, we envisioned to enhance with an AI engine that helps the evaluation of the existing components.
In particular, when we consider existent softwares that use machine learning techniques to score or dock a ligand \cite{isert2023structure}, they use experimental data as the training set. These training sets are limited in number and include only \emph{positive samples}.
Therefore, by using the designed CADD workflow we can increment the sheer number of available data by generating them in-silico.
This solution based on the in-silico free energy calculation can generate also data for negative samples since we can feed the module with ligands and poses that cannot occur naturally.

\section{Main Goals and Achievements} 

This section briefly describes the main goals of the LIGATE project and provides an outlook on the main achievement reached so far in the project.

\subsection{Performance Portability} 
Modern high-performance computing (HPC) systems leverage hardware heterogeneity (e.g. GPUs, FPGAs) as a key feature. 
In the HPC scenario, the Top500 list \footnote{Top500 List Website - \url{https://www.top500.org/}} shows that, at the end of 2022, eight out of the ten top systems are GPU-accelerated, with NVIDIA and AMD being the primary hardware providers. 
However, applications developed for GPUs are often written in proprietary, non-portable programming languages, such as CUDA for NVIDIA GPUs, which can lead to vendor lock-in and hinder the ability to quickly migrate to different platforms. Moreover, being bound to a single architecture can significantly impact the application's ability to be applied in urgent computing scenarios where the maximum computing power needs to be harnessed. Therefore, it is crucial to have application code that can efficiently run on different devices. 

In recent years, several programming models for heterogeneous computing have been proposed. One such model is SYCL\cite{SYCL2020}, a royalty-free, cross-platform abstraction layer that enables code for heterogeneous processors to be written using standard ISO C++. SYCL allows for both the host and kernel code for an application to be in the same source file. The Kronos Group consortium maintains SYCL, and it has multiple implementations provided by companies, such as Intel \cite{oneapi}, and universities \cite{opensycl}. SYCL enables the development of high-level, performance-portable applications that can target CPUs, GPUs, and FPGA cards without requiring device-specific code.

Within LIGATE, we designed a \textit{performance-portable} implementation of LiGen, originally written in C++ and CUDA to target only NVIDIA GPUs \cite{9817028, vitali2019exploiting}. We ported the legacy code to SYCL2020 analyzing different alternative possibilities (e.g., adopting Unified Shared Memory and Buffer-Accessors paradigms). The new SYCL implementation can target CPUs, as well as both NVIDIA and AMD GPUs, unlocking the possibility to run on the GPU partition of LUMI supercomputer and on an E4 internal research cluster.

\begin{figure}[t]
    \centering
    \includegraphics[width=\linewidth]{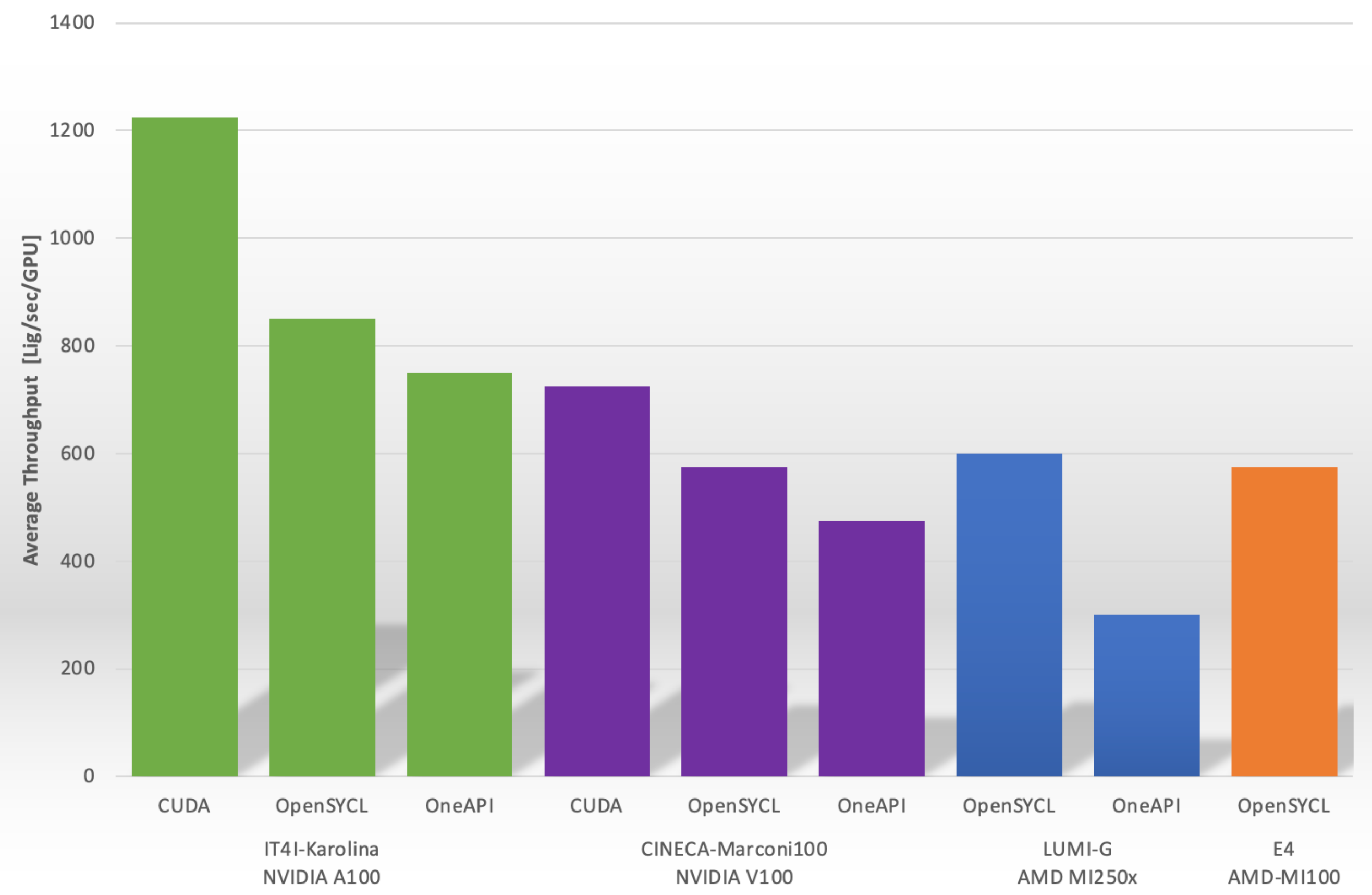}
    \caption{LiGen Performance Results on different Clusters (IT4I-Karolina, CINECA-Marconi100, LUMI-G and E4), and using different implementations (CUDA and SYCL)and software stacks (OpenSYCL, OneAPI). }
    \label{fig:LigBench}
\end{figure}

Figure \ref{fig:LigBench} presents the performance results in terms of Ligands per second per GPU  with both the native CUDA application and the SYCL version. While the former can only run on NVIDIA hardware, for the SYCL version we show the results on both NVIDIA (A100, V100) and AMD (MI250X, MI100) GPUs using the OpenAPI and OpenSYCL compilers. The results have been benchmarked on three top European supercomputers, and on the E4 research cluster.
While the native CUDA version achieves a higher throughput on the same hardware, the SYCL version is able to run also on AMD GPUs with comparable performance, while also proving a ground base for following optimizations tailored to novel systems (e.g. LUMI cluster).

\subsection{The Celerity Runtime System}
While SYCL aids in alleviating the portability issues of single-GPU code across vendors, developers of applications targeting a distributed memory cluster with accelerators still have to contend with manually implementing communication. Doing so efficiently, especially in the presence of the large amount of low-level details introduced by GPU computing and network communication across different vendors, topologies, and technology stacks, is a challenge even for parallel system experts. 
The Celerity Runtime System~\cite{salzmann2023async} provides a high-level C++ layer closely inspired by SYCL, which automatically manages work distribution and data dependencies. It enables high-productivity development of applications for GPU clusters~\cite{thoman2019celerity}, only requiring efforts comparable to a baseline single-GPU SYCL implementation.

Within the LIGATE project, we have extended the Celerity system with new high-level primitives to more directly encode the semantics of the target applications~\cite{thoman2022celerity}, and we improved the throughput on large systems by introducing a zero-communication-overhead distributed command generation scheme~\cite{salzmann2023async}.

\subsection{LiGen Autotuning}

In the context of HPC, autotuning \cite{autotuningHPC, antarex} refers to the process of automatically selecting the best configuration of the software or compiler parameters according to a specific target hardware and execution condition. This automatic process permits a more efficient execution of the code eliminating the need for manual intervention. 
The importance of autotuning is not only for a specific application-architecture pair, but it has a huge impact when it is important to guarantee performance portability across multiple architectures or application versions that can exploit a different best configuration. 

Several autotuning software solutions exist \cite{ansel2014opentuner, gadioli} that allow to efficiently explore the configuration space with varying degrees of complexity, with or without the possibility of taking constraints into account. 
Within LIGATE, our intention has been to exploit this possibility together with an explicit parameterization of one of the main LIGATE components.
In particular, we revised the original LiGen code by exposing as much as possible the internal parameters that can have an impact on both quality and performance, or that can spot different optimization opportunities on different architectures. 
We extracted 11 software parameters (knobs) of the LiGen application, accounting for overall more than 60 million different configurations, and we run a tuning campaign to search for Quality-Performance tradeoffs.
The importance of the trade-off is the possibility to use the same docking tool for both small-scale campaigns requiring higher quality, and extreme-scale virtual screening experiments where the overall duration of the campaign and resource usage (cost) can be the actual limit.
To make the exploration feasible from the execution time perspective, we adopted an iterative approach based on the Bayesian Optimization (BO) algorithm supported by Machine Learning techniques, named MALIBOO \cite{guindani2022maliboo}. This approach is efficient in minimizing the execution costs of recurring resource-constrained computing jobs.

\begin{figure}[t]
    \centering
    \includegraphics[width=\linewidth]{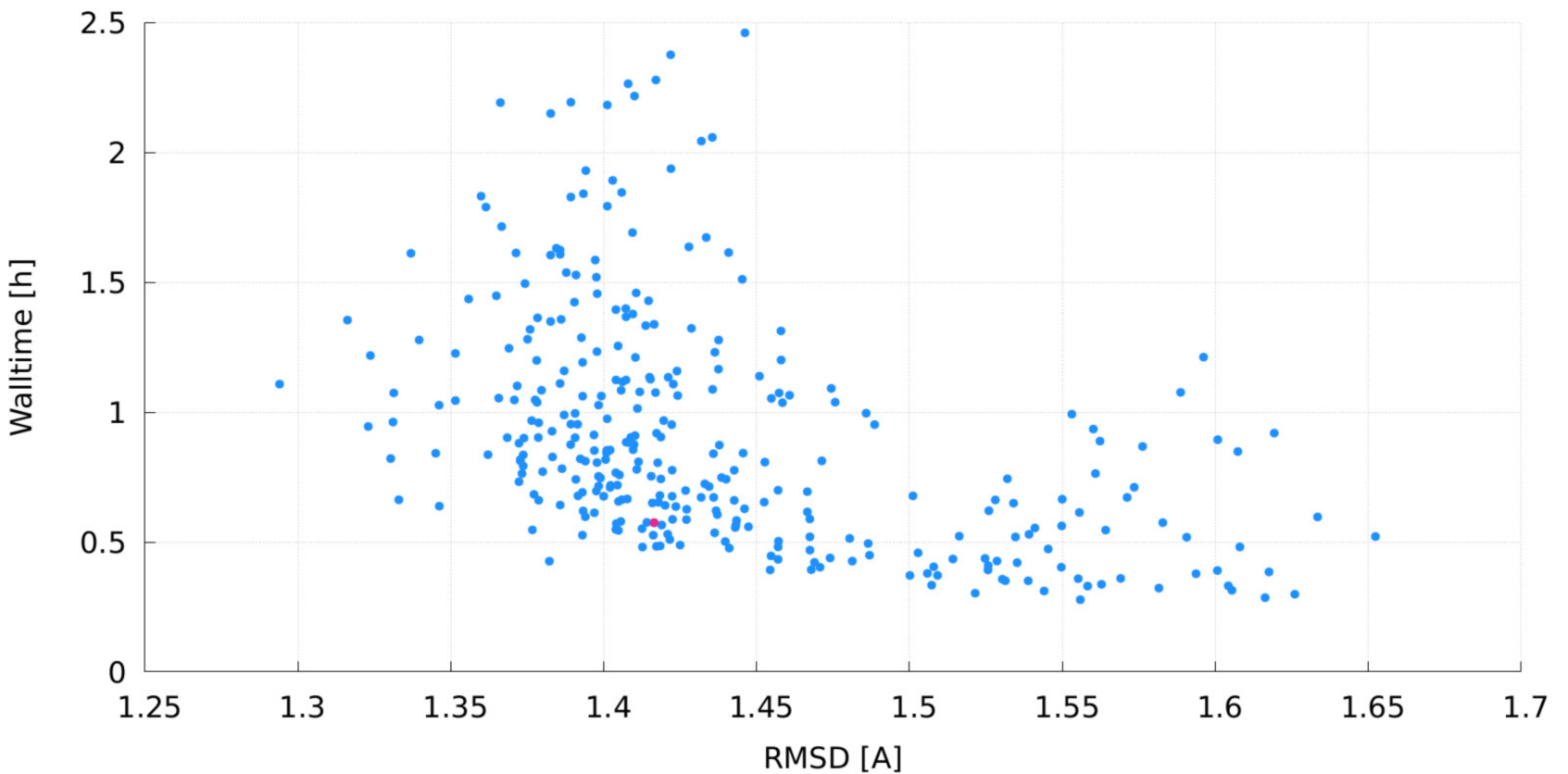}
    \caption{Exploration of the LiGen software knob}
    \label{fig:LigEXPLORE}
\end{figure}

Figure \ref{fig:LigEXPLORE} shows an example of the exploration results where obtained evaluating a small subset of 320 different LiGen configurations (blue dots). In particular, the RMSD (Root Mean Square Distance) calculation refers to the distance between the predicted best pose and the actual experimental pose for each ligand-protein pair derived from the CASF2016 dataset \cite{CASF16}. The wall-time refers to the overall time needed to screen 1M ligands.


In addition to the parameters that are explicitly exposed by LiGen, we also exploited the possibility of auto-tune part of the code according to the characteristics of the input data, i.e. the ligands.
In fact, to target high-throughput experiments making better use of GPUs, we completely re-designed LiGen in a way that it processes a \emph{batch} of ligand-protein pairs across the device, instead of a single one as done by state-of-the-art approaches \cite{10.48550/arxiv.2209.05069}. 
To maximize the efficiency of the new approach the ligands are clustered by size before the GPU kernels and are temporarily stored in a set of buffers. The GPU computation is launched once one of the batches reached a specific size that is dependent on the ligand characteristics, i.e. number of atoms and rotatable bonds, and GPU memory/micro-architecture.
The fine-grain tuning of the batch size for each ligand size has been derived using an abstract model of the device that resulted to be portable across the different analyzed GPUs.


\subsection{Integrated automatic setup and execution of free energy calculations}

The main goal of the molecular simulation work in LIGATE is to provide
a high-performance implementation of accurate free energy calculations that is
also fully automated and portable. To achieve this, we have first developed a workflow ranking drug candidates by their relative free energy of binding (RBFE) to the target protein. Starting from lists of thousands of docked compounds/poses, this workflow with its new interfaces can automatically determine the force field parameters for both the protein and the compound, set up free energy calculations in GROMACS, execute the entire ensemble as a parallel run on HPC resources using HyperQueue (See Section 3.6), and return free energies with a requested standard error. 

To obtain the RBFE estimates with a minimal number of molecular dynamics (MD) steps, the compounds are grouped into pairs that are alchemically transformed into each other during the MD simulations, which requires the workflow to solve the maximum common substructure problem for each compound pair. In the MD simulations, the Accelerated Weight Histogram method (AWH) creates a bias potential to facilitate the alchemical transformation of the compound pair, and this bias potential converges to the RBFE.~\cite{Lundborg2021}. For GROMACS-2023, the performance of free energy calculations has also been improved by rewriting the 
free energy non-bonded kernels to use SIMD instructions, which more than doubles
performance when the algorithm is a bottleneck, which was typically the case when running e.g. on GPU accelerators.

Finally, to build an AI engine supporting the analysis in the early stages of the CADD platform, it is necessary to go beyond relative free energies and rather target 
absolute binding free energies (ABFEs) of protein-compound complexes. To this end, we are developing a variant of the above workflow in which the free energy calculations are replaced with standard MD simulations with GROMACS. ABFEs are subsequently estimated from the MD trajectories as the sum of gas and solvation free energy with gmx\_MMPBSA.~\cite{ValdesTresanco2021}


\subsection{Machine Learning engines to further speedup the virtual screening phase}
In LIGATE, we want to exploit the possibility of generating a large set of data to analyze the possibility to use Machine Learning (ML) techniques to further speed up the virtual screening phase. In particular, one of the goals of the project is to evaluate the possibility of predicting the binding affinity in the early stages of the CADD platform, using the costly free energy computation derived by GROMACS as ground truth. 

Since the design of the best prediction algorithm for a given class of applications strictly depends on the problem at hand and on the available data, this task will evaluate different machine learning approaches starting from simple, computationally efficient techniques (e.g. multivariate regressions) up to techniques demanding more computational effort, such as deep/graph neural networks. 
The most suitable prediction technique for the CADD platform has to be identified considering the cost to obtain the prediction in relation to the specific usage of the CADD platform. 
Fast screening of a large ligand database (e.g. in case of urgent computing) would require a cheaper prediction model, while a more detailed analysis on a reduced set of molecules can use a more computationally demanding model. Starting from the scoring function already available in LiGen \cite{beccari2013ligen}, we will identify the features related to ligand, pocket and their interaction to be used as input variables of the free-energy prediction model to be used. Moreover, we are going to analyze the possible subset of them that can be used also to steer the pose generation during the geometric docking phase \cite{gadioli2021tunable} which is the core engine within LiGen.

While waiting for the data to be generated by free energy calculations on GROMACS, we analyzed the problem of pose selection. Pose selection is the step within the structure-based virtual screening phase that is in charge of the selection of the most reliable and interesting poses to be evaluated in terms of binding affinity. We generated a large set of data using LiGen and we labeled them using the RMSD with respect to known experimental poses. Several state-of-the-art machine learning methods have been analyzed while looking at the most promising ones. 3D-CNN (Convolutional Neural Networks, i.e. Pafnucy \cite{3DCNN}, and DimeNet \cite{dimenet}), GNN (Graph Neural Networks, i.e. GraphBar \cite{graphbar}), and MDN (Mixture Density Network, i.e. DeepDock \cite{DeepDock}, and RTMScore \cite{rtmscore}) have been retrained using the generated large set of RMSD labeled data and are currently under investigation for a seamless integration within the pipeline. 

Moreover, a novel pose selector scheme called LiHPS (LiGen High-throughput Pose Selector), targeting extreme-scale virtual screening campaigns, has been developed from scratch adopting a mix of approximate computing, and supervised/unsupervised techniques. With respect to the previously mentioned 3D-CNN/GNN/MDN methods, LiHPS exploits another sweet spot in the throughput-accuracy trade-off curve, where a small reduction in accuracy brings a very large performance improvement.

\subsection{Efficient sub-node task scheduler.} Modern HPC clusters contain a large number of heterogeneous resources that provide vast amounts of computational power. The potential of these clusters might not be fully exploited when users submit a large number of tasks that cannot utilize the whole computational node. Existing HPC job managers are not well prepared for such scenarios, which can lead to nodes being underutilized.

HyperQueue is a task runtime framework \footnote{https://github.com/It4innovations/hyperqueue} designed for the efficient execution of a large number of heterogeneous tasks on HPC systems, which was developed within LIGATE. It takes care of asking for computational resources from the job manager on behalf of the user, and load balances submitted tasks on all available computing nodes with a sophisticated scheduler to keep node utilization as high as possible.

HyperQueue has been used within LIGATE to simplify the execution of the target in-silico virtual screening application, which is composed of many small tasks. Using HyperQueue, users of the application do not have to care about splitting the tasks into individual HPC jobs and then submitting these jobs separately. They simply submit all tasks into HyperQueue and it takes care of the rest.

\subsection{Virtual Screening-as-a-Service.}
Among the several goals, LIGATE directly addresses the contemporary challenges that running a virtual screening campaign at a large scale requires extensive expertise and manual intervention. This is motivated by the fact that (i) there is a lack of ability for different tools to interoperate efficiently and automatically through APIs and well-specified data formats, and (ii) that the efficient management of large HPC resources requires additional skills that are not always available for domain experts.

To address this issue, we integrated the LiGen molecular docking application with an HPC-as-a-Service framework (HEAppE \footnote{HEAppE Website - \url{https://heappe.eu/web}}) \cite{Svaton2020280} to permit the execution of virtual screening tasks on supercomputing resources through REST API without a direct access to them. This allowed us the simple integration with the LEXIS Platform \cite{Golasowski202217} which provides an easy to use web-portal to manage data and workflow execution on remote HPC resources.

The LEXIS platform and the integrated LiGen-related workflow have been designed to enable multi-site execution of a large-scale computation in case of urgent computing, thus permitting the deployment of part of the virtual screening campaign on multiple HPC centers without the need for explicit management. A central part of this platform is the support for distributed data management between different sites together with data staging to/from HPC resources. 
The LEXIS-LiGen integration permits also to envision the use of the LIGATE platform at the end of the project in a Virtual Screening-as-a-service fashion for researchers and public institutions.
\subsection{Scientific validation of the CADD platform}

The LIGATE CADD platform is a promising tool for structure-based drug discovery due to its key characteristics: portability, tunability, and scalability. 
 However, we plan to demonstrate the effectiveness of the technology not only by considering extra-functional aspects but also analyzing the actual advantages that the innovative CADD platform can bring to the target drug discovery field.

To do so, we have compiled comprehensive datasets of protein-ligand complexes that encompass a diverse range of drug targets and small molecules. These benchmark datasets are derived from the PDB and meet rigorous quality criteria, serving as a reliable "ground truth" for the evaluation of the LIGATE CADD platform. We have ensured the diversity of the datasets from three perspectives: protein, ligand, and binding pocket, ensuring sufficient representation of each. Moreover, we have included special sets of mutant structures and similar ligands to better assess the platform's performance in challenging cases. We then developed an automated pipeline for continuous benchmarking and evaluation of the virtual screening platform, with automated reporting on the scale and areas of improvement.

As the final step, the project consortium plan to select antimicrobial drug targets for infectious diseases as real-world test cases. \textit{In vitro} experimental validation of hits identified by the CADD platform would demonstrate the practical applicability of the approach on disease targets with high medical relevance according to the WHO priorities. 
The results will be published in peer-reviewed journals, and benchmark data will be made openly accessible under FAIR data principles.


\subsection{Key Technologies on the Cronos Use Case}
To further validate the technology developed to support the tuning, parallelization, and portability of the CADD platform for extreme-scale virtual screening, within the project an additional use case has been considered. 
The selected application, Cronos~\cite{kissmann2018cronos}, comes from the astro- and particle physics world, a completely different field with respect to LiGen and GROMACS. A structured grid magneto-hydrodynamics simulation in 3D that resolves plasma-dynamical problems in astrophysics and space science. 
%
The original implementation is written in C++, and parallelized using MPI to target CPU-based distributed memory systems. 
This version, while demonstrating excellent scalability on medium-sized clusters (approx.\ 95\% parallel efficiency on 1024 cores, 60-70\% on 16,384 cores), does not use GPUs despite the fact that structured grid applications are often well-suited for accelerated clusters.

The goal for Cronos within LIGATE is (1) to have a SYCL and Celerity~\cite{salzmann2023async} implementation to enable a portable GPU execution, (2) to rigorously test both platform suitability for this use case as well as performance and scalability, and (3) to investigate and benchmark possible performance optimizations.
At the current stage of the project, a functional SYCL porting is available. A deviation in the output from the CPU to GPU execution originates from the different implementations of reduction algorithms coupled with floating-point math semantics. The Celerity implementation is ongoing, after which initial performance tests can be performed to drive further optimization work.

\section{Conclusions}

Although capable of accurate predictions, \textit{in-silico} tools like virtual screens with docking and free energy calculations used to be limited by the availability of experimental structures of the target protein. With the accurate protein structure predictions provided by AlphaFold2,~\cite{Jumper2021} starting structures for the vast majority of target proteins can now be generated computationally; and thanks to the large chemical space covered by the LiGen tools, protein-compound complex structures can be obtained entirely \textit{in silico} for a very diverse and chemically motivated set of drug candidates. 
The integration of docking and free energies from MD in the CADD workflow we present here is thus timely to exploit the new potential of these \textit{in-silico} tools and foster methodological developments; in particular, free energy labels from MD for docking poses can help to improve the scoring functions used in docking via AI engines trained on these MD free energy labels.

Complementing the increased scientific scope of application of docking and free energy calculations, performance portability of the respective software packages, LiGen and GROMACS, and efficient task scheduling with HyperQueue ensure their wide technical applicability, and the vast computational resources currently deployed e.g. in EuroHPC will open new opportunities to generate synthetic training data for machine learning
in drug design applications.

Overall, the LIGATE project aims to provide an automated solution for CADD on EuroHPC resources from the current pre-exascale architectures to the future exascale systems. The integration of best-in-class European software components will be made available to industrial as well as non-profit academic research and public institutions, as a Platform-as-a-Service solution for drug discovery.


\section*{Acknowledgements}
This project has received funding from EuroHPC-JU - the European High-Performance Computing Joint Undertaking - under grant agreement No 956137. The JU receives support from the European Union’s Horizon 2020 research and innovation programme and Italy, Sweden, Austria, Czech Republic, and Switzerland. We acknowledge EuroHPC Joint Undertaking for awarding us access to Karolina at IT4Innovations, Czech Republic (EHPC-DEV-2021D02-049), and to LUMI, hosted by CSC (Finland) and the LUMI consortium (EHPC-BEN-2022B12-001). Additional computational resources were provided by the Swedish National Infrastructure for Computing grant 2022-3/40.

\bibliographystyle{ACM-Reference-Format}
\bibliography{sample-base}
\end{document}